\documentclass[nocopyrightspace]{sig-alternate-05-2015}
\makeatletter
\def\@copyrightspace{
	\@float{copyrightbox}[b]
	\begin{center}
		\setlength{\unitlength}{0.2pc}
		\begin{picture}(100,0) %Space for copyright notice
		\put(0,-0.95){\crnotice{\@toappear}}
		\end{picture}
	\end{center}
	\end@float}
\makeatother

\usepackage{amsthm}
\newtheoremstyle{example}% name
{9pt}%      Space above, empty = `usual value'
{9pt}%      Space below
{\itshape}% Body font
{}%         Indent amount (empty = no indent, \parindent = para indent)
{\bfseries}% Thm head font
{:}%        Punctuation after thm head
{ }% Space after thm head: \newline = linebreak
{}%         
\theoremstyle{example} 
\newtheorem{eg}{Example}

\newtheoremstyle{definition}% name
{9pt}%      Space above, empty = `usual value'
{9pt}%      Space below
{\itshape}% Body font
{}%         Indent amount (empty = no indent, \parindent = para indent)
{\bfseries}% Thm head font
{:}%        Punctuation after thm head
{ }% Space after thm head: \newline = linebreak
{}%         
\theoremstyle{definition} 
\newtheorem{defn}{Definition}

\newtheoremstyle{definition}% name
{9pt}%      Space above, empty = `usual value'
{9pt}%      Space below
{\itshape}% Body font
{}%         Indent amount (empty = no indent, \parindent = para indent)
{\bfseries}% Thm head font
{:}%        Punctuation after thm head
{ }% Space after thm head: \newline = linebreak
{}%         
\theoremstyle{definition}

\usepackage{cite}
\usepackage{amsmath,amssymb,amsfonts}
\usepackage[linesnumbered,ruled,vlined]{algorithm2e}
\usepackage{graphicx}
\usepackage[export]{adjustbox}
\usepackage{textcomp}
\usepackage{xcolor}
\usepackage{amsthm}
\usepackage{subcaption}
\usepackage{floatrow}
\setcopyright{none}

\begin{document}

\title{SlotSwapper: A Schedule Randomization protocol for Real-Time WirelessHART Networks}
%\subtitle{[Extended Abstract]
%\titlenote{A full version of this paper is available as
%\textit{Author's Guide to Preparing ACM SIG Proceedings Using
%\LaTeX$2_\epsilon$\ and BibTeX} at

\author{\alignauthor Ankita Samaddar, Arvind Easwaran, Rui Tan\\
% 1st. author
       \affaddr{Nanyang Technological University, Singapore}\\
       \email{\{ankita003,arvinde,tanrui\}@ntu.edu.sg}
}

\toappear{\hrule \vspace{0.1cm} \tiny {\bf \textcopyright2019 Copyright held by authors.}\\ \tiny \bf{RTN'19, July 2019, Stuttgart, Germany}}
\maketitle{}

\begin{abstract}
Industrial process control systems are time-critical systems where reliable communications between sensors and actuators need to be guaranteed within strict deadlines to maintain safe operation of all the components of the system. WirelessHART is the most widely adopted standard which serves as the medium of communication in industrial setups due to its support for Time Division Multiple Access (TDMA) based communication, multiple channels, channel hopping, centralized architecture, redundant routes and avoidance of spatial re-use of channels. However, the communication schedule in WirelessHART network is decided by a centralized network manager at the time of network initialization and the same communication schedule repeats every hyper-period. Due to predictability in the time slots of the communication schedule, these systems are vulnerable to timing attacks which eventually can disrupt the safety of the system. In this work, we present a moving target defense mechanism, the SlotSwapper, which uses schedule randomization techniques to randomize the time slots over a hyper-period schedule, while still preserving all the feasibility constraints of a real-time WirelessHART network and makes the schedule uncertain every hyper-period. We tested the feasibility of the generated schedules on random topologies with 100 simulated motes in Cooja simulator. We use schedule entropy to measure the confidentiality of our algorithm in terms of randomness in the time slots of the generated schedules. 
\end{abstract}

%\begin{CCSXML}
%<ccs2012>
%  <concept_id>10010520.10010553.10010562</concept_id>
%  <concept_desc>Computer systems organization~Embedded systems</concept_desc>
%  <concept_significance>500</concept_significance>
%  <concept_id>10010520.10010575.10010755</concept_id>
%  <concept_desc>Computer systems organization~Redundancy</concept_desc>
%  <concept_significance>300</concept_significance>
%  <concept_id>10010520.10010553.10010554</concept_id>
%  <concept_desc>Computer systems organization~Robotics</concept_desc>
%  <concept_significance>100</concept_significance>
%  <concept_id>10003033.10003083.10003095</concept_id>
%  <concept_desc>Networks~Network reliability</concept_desc>
%  <concept_significance>100</concept_significance>
%</ccs2012>  
%\end{CCSXML}
%
%\ccsdesc[500]{Computer systems organization~Embedded systems}


%
%  Use this command to print the description
%
%\printccsdesc

% We no longer use \terms command

\section{Introduction}\label{sec:intro}

Time-critical systems such as the industrial process control systems are real-time cyber-physical systems (CPS) that monitor and control the production lines in a manufacturing plant. The number of devices in such setup keeps increasing. To support more devices and to cope up with frequent changes in the network topology due to addition (removal) of devices to (from) the network, a switch of the communication infrastructure from wired networks to wireless networks is desirable. Among the existing wireless sensor network (WSN) standards, WirelessHART is best suited for the industrial process control systems due to its reliable TDMA-based schedule, centralized architecture, multi-channel support, channel hopping, redundancy in routes, and avoidance of spatial re-use of channels.

%Cyber-physical systems (CPS) cover a large spectrum of today's life. A large class of CPS are governed by strict timing requirements and are referred to as real-time CPS. One such real-time CPS is the industrial process control system that monitors and controls the production lines in the manufacturing plants. Rapid advancement of technology and application of automation in these industrial setups have resulted in constant increase in the number of devices. To support more devices and to cope up with frequent changes in the network topology due to addition (removal) of devices to (from) the network, the communication infrastructure in these setups are switching from wired to wireless connection. Among the existing wireless sensor network (WSN) standards, WirelessHART is best suited for the industrial process control systems due to its reliable TDMA-based schedule, centralized architecture, multi-channel support, channel hopping, redundancy in routes, and avoidance of spatial re-use of channels. As a result, WirelessHART standard has been widely adopted in industrial setups. 
%As of 2012, it had been adopted in over 8,000 manufacturing systems~\cite{wirelesshart-deploy}. 
Although the use of wireless brings flexibility and adaptability to the communication infrastructure, it increases the threats of cyber attacks. Some recent sophisticated attacks against critical infrastructures such as Stuxnet~\cite{langner2011stuxnet} and Dragonfly~\cite{dragonfly} have alerted us to the shaky protection of the conventional air gap solution. 
%Though WirelessHART is adopted in industrial process control systems for its reliability, it is not fully secure against cyber attacks. 
The main components of a WirelessHART network are the sensors, actuators, Gateway, a network manager, and multiple access points~(AP). Each communication between these devices are real-time flows with fixed periods and deadlines. To make the flows schedulable, the schedule in a WirelessHART network is pre-determined by the centralized network manager at the time of network initialization. The same schedule is repeated over every hyper-period ({\em i.e.}, lowest common multiple of the periods of all the flows in the network), until there is any change in the network topology, such as addition/removal of new/existing devices to/from the network. The repetitive execution of the deterministic flow schedule in a WirelessHART network over every hyper-period makes these systems vulnerable to timing attacks. Such repetition greatly helps the attacker to analyze the eavesdropped traces and infer the schedule. With the inferred schedule, the attacker can further launch various strategic destructive attack steps. For instance, the attacker can selectively jam the transmissions from/to a certain critical sensor/actuator which can eventually breach the safety of the system. 

%In particular, the wireless communications in industrial settings often follow regular patterns, such as the repetitive executions of the deterministic flow schedule of a WirelessHART network every hyper-period.
%The attacker may use insiders to deploy certain attack devices with radio frequency (RF) interfaces on the factory floors or compromise existing wireless devices. The broadcast nature of wireless communications makes it easy for the attacker to monitor and interfere with the legitimate transmissions.
%Various cyber-physical systems (CPS) impose real-time requirements. One such real-time CPS is the industrial process control system that monitors and controls the production lines in a manufacturing plant. The number of devices in such industrial CPS keeps increasing. To support more devices and to cope up with frequent changes in the network topology due to addition (removal) of devices to (from) the network, a switch of the communication infrastructure from wired networks to wireless networks is desirable. Among the existing wireless sensor network (WSN) standards, WirelessHART is best suited for the industrial process control systems due to its reliable TDMA-based schedule, centralized architecture, multi-channel support, channel hopping, redundancy in routes, and avoidance of spatial re-use of channels. As a result, WirelessHART standard has been widely adopted in industrial setups. As of 2012, it had been adopted in over 8,000 manufacturing systems \cite{wirelesshart-deploy}.
In this work, we aim at reducing the predictability of the time slots in the communication schedule of a real-time WirelessHART network. We propose a moving target defense~(MTD) mechanism, the {\em SlotSwapper}, that randomizes the time slots in the communication schedule over every  hyper-period, satisfying all the feasibility constraints of a real-time WirelessHART network as follows--- (1) deadlines of all real-time flows in the network are to be satisfied, (2) the hop sequences associated with each flow are to be preserved and (3) no conflicting transmissions in the network are allowed. From our analysis, the attacker who can monitor the wireless transmissions needs at least two hyper-periods to infer the schedule. Randomizing the schedule over every hyper-period renders the attacker's inference futile, thereby greatly improving the confidentiality of the WirelessHART network's operations. 
%The {\em SlotSwapper}, consists of two main phases--- (1) an offline schedule generation phase in which it generates a set of hyper-period schedules by randomly swapping the flow instances in the slots without violating the three feasibility constraints of a real-time WirelessHART network (2) an online schedule selection phase during which all the nodes select the same schedule uniformly at random every hyper-period from the set of generated schedules and follow the schedule over that hyper-period. 
More varied are the slots in a schedule, more difficult it is for the attacker to predict them. Hence, the measure of uncertainty in the time slots of a schedule can be expressed in terms of the amount of randomness in the time slots over the hyper-period schedules generated by our algorithm. We re-defined {\em schedule entropy}~\cite{yoon2016taskshuffler} as a metric to measure the uncertainty in predicting the time slots. 
%and the Kullback-Leibler divergence~\cite{raiber2017kullback} as a metric to measure the divergence of our MTD solution with respect to a hypothetical truly random solution. 
We illustrated the feasibility of our proposed algorithm on random topologies with 100 simulated nodes in Contiki Cooja~\cite{osterlind2006cross}. To the best of our knowledge, this is the first work on randomization to reduce the determinism of the time slots of a hyper-period schedule in real-time WirelessHART networks.

\section{Related Work}\label{sec:related}
%Two major challenges in real-time systems are reliability and security--- achieving both of them at the same time is quite difficult. 
Two notable works in the literature which adopt randomization techniques in the context of real-time processor scheduling are {\em taskshuffler}~\cite{yoon2016taskshuffler} and SPARTA~\cite{jiang2016sparta}. 
%exist in the literature in the context of real-time systems security in which countermeasures to attacks are provided by adopting certain techniques in the scheduling policy of the system. 
\cite{yoon2016taskshuffler} presents a schedule randomization protocol, the {\em taskshuffler}, that shuffles a set of fixed priority real-time tasks on a uniprocessor system. 
%Security in real-time systems is very hard to achieve because of strict deadline constraints of the real-time tasks in these systems. A number of works exist in the literature in the context of real-time systems security. \cite{yoon2016taskshuffler} presents a schedule randomization protocol, the taskshuffler, which shuffles a set of fixed priority real-time tasks on a uniprocessor system to make the schedule of communication unpredictable to an attacker. %\cite{veyrat2012shuffling} proposes two types of side channel attack (SDA) causing information leakage thereby breaching the {\em confidentiality} of the system. They propose a solution to this problem by randomizing the execution order of both instructions and resources which in turn preserves the confidentiality of the system.
%\cite{li2011hijacking} describes both active and passive attacks on an insulin pump by introducing a Universal Software Radio Peripheral (USRP). As a remedy to this problem, the system proposes some traditional cryptographic approach and a body coupled communication network to provide safety to the system. 
\cite{jiang2016sparta} proposes SPARTA, a scheduler to randomize the leakage points in the schedule protecting the system from Differential Power Analysis (DPA) attacks. However, both of these works are on uniprocessor system. Our problem is even harder than multi-processor scheduling. $m$ channels and $n$ real-time flows of our network can be mapped to $m$ processors and $n$ real-time tasks respectively. However, the conflicting transmissions among the flows impose additional constraint in our network which makes our problem even harder than multi-processor scheduling.

Due to support for TDMA schedule in WirelessHART networks, these networks are vulnerable to selective jamming attacks~\cite{proano2010selective}. \cite{mpitziopoulos2009survey,virmani2014routing} survey various possible jamming attacks and the key ideas of existing security mechanisms against such attacks in WSNs. 
%\cite{xu2005feasibility} classifies the jamming attacks in WSNs into four categories, {\em constant}, {\em deceptive}, {\em random}, and {\em reactive} jamming. 
\cite{pongaliur2008securing} proposes various types of side-channel attacks and their respective countermeasures in WSN. The countermeasures against jamming attacks can be provided from physical-layer solutions as in \cite{mpitziopoulos2009survey,pickholtz1982theory} or cyber-space solutions such as \cite{proano2012packet,wood2007deejam}. 
%It also provides a countermeasure of encrypting the packets using cryptographic primitives. 
%\cite{daidone2014solution} provides a centralized solution, in which the WSN nodes regularly exchange messages with a coordinator node. 
\cite{stojanovski2015efficient} presents the steps of an attacker to launch jamming attacks in industrial process control systems. Recent works such as \cite{tiloca2017jammy} and \cite{tiloca2018dish} provide countermeasures against timing attacks in single and multi-channel WSN respectively by permuting the slot utilization pattern at the node level over a super-frame to randomize the schedule. However, the flows considered in these works are not associated with deadlines, hence, randomization of slot utilization pattern at the node level makes the flows schedulable. Our problem is more complex. Each flow in our network is a real-time flow with a strict deadline. Permuting the time-slots at each node does not guarantee deadline satisfaction of all the real-time flows in our network, hence, existing solutions in \cite{tiloca2017jammy} and \cite{tiloca2018dish} are not applicable.
%\cite{rahbar2006analysis,jang2016breaking,marco2014effectiveness} discuss about address space layout randomization (ASLR) in memory.
%Several studies also exist in the context of address space layout randomization (ASLR) in memory  which randomize the memory locations to reduce the predictability of the memory accesses. 
%In this paper, we present a threat model and an MTD mechanism to mitigate the threat. Our countermeasure randomizes the time slots in a schedule such that the predictability of the scheduled time slots is reduced significantly.
%We also provide countermeasure to the attack by incorporating randomization techniques in the schedule to obfuscate the determinism in the time slots of the schedule.
%attack model in real-time WirelessHART network where the communication schedule is pre-decided by the network manager. Hence, the attacker can target specific links in the network and launch timing attacks into the system. 
%Some existing works in the context of security in wireless sensor networks (WSN) are also highlighted in this section.   This work proposes a symmetric key cryptography as a solution to the problem. \cite{mpitziopoulos2009survey},\cite{virmani2014routing} are surveys on the different types of jamming attacks in WSN and discusses different countermeasures that can be adopted to avoid those attacks. 
\section{WirelessHART Background}\label{sec:background}

The WirelessHART protocol, being compliant with IEEE 802.15.4, is the first open wireless communication standard for measurement and control in network and process industry~\cite{lu2016real}. A WirelessHART network 
%to provide reliable digital communications to satisfy stringent deadline requirements of industrial applications. 
consists of a Gateway, multiple field devices, APs and a centralized network manager which are connected via wireless mesh networks. The network manager, connected to the Gateway, is responsible for managing the devices, scheduling, creating the routes and optimizing the network. The field devices are wireless sensors and actuators which can either transmit or receive in a particular time slot. Also, in a time slot, a receiver can receive from exactly one sender. Multiple APs are connected to the Gateway via wired connections to provide redundant paths between the Gateway and the network devices. The key features of the WirelessHART network for which it is suitable for process industries include

\vspace{0.5em}\noindent \textbf{TDMA:} For reliable collision-free communications in a WirelessHART network, time is globally synchronized and slotted into $10$ms time slots within which a network device sends a packet and receives its corresponding acknowledgment.   
	%In WirelessHART network, time is globally synchronized and slotted into time slots of $10ms$  This results in predictable communication latencies, thereby ensuring deadline satisfaction of the real-time flows.
	%A time slot can be either a dedicated slot or a shared slot. In a dedicated time slot, only one sender is allowed to transmit a packet to its intended receiver. In a shared slot, more than one sender can attempt to transmit to the same receiver. In order to reduce collisions, the senders adopt carrier sense multiple access with collision avoidance (CSMA/CA) to contend in a shared slot.

%\vspace{0.5em}\noindent \textbf{Size of the network:} 
%Deployment of a large-scale Wireless Sensor Actuator Network (WSAN) is extremely challenging. 
%A typical WirelessHART network consists of approximately $100$ field devices under the control of a single Gateway. %which is sufficient to manage the system from the centralized manager.
%Large-scale WSANs can be formed by initiating communications between the Gateways. 
%In a typical WirelessHART network, the number of hops between the Gateway and the source or destination node is no more than $4$  hops~\cite{alur2009modeling}.
	%The number of hops between the devices and the Gateway is at-most four~\cite{alur2009modeling}. A network size of $100$ devices   
\vspace{0.5em}\noindent \textbf{Channel and route diversity:} WirelessHART supports a maximum of $16$ channels~\cite{Chen:2010:WRM:1855162} at a frequency band of 2.4~GHz. To avoid interference from neighboring wireless systems, it adopts channel hopping in every time slot. A channel is blacklisted if it suffers from external interference. WirelessHART allows route diversity by transmitting a packet multiple times via multiple paths over different channels. 
%The entire frequency band of WirelessHART is divided into $16$ channels~\cite{Chen:2010:WRM:1855162}. To avoid interference from neighboring wireless systems, it adopts channel hopping in every time slot. Also, if any channel suffers from external interference, that channel is blacklisted and not used for communication.
%\vspace{0.5em}\noindent \textbf{Route diversity:} WirelessHART allows route diversity by transmitting a packet multiple times via multiple paths over different channels. 

\vspace{0.5em}\noindent \textbf{Avoidance of spatial re-use of channels:} 
%Due to variability in the interference patterns between the nodes, the interference in WSN is very hard to detect. 
To avoid interference and to increase reliability, WirelessHART avoids spatial re-use of channels~\cite{Chen:2010:WRM:1855162}.
%{\em i.e.}, in a time slot only one communication occurs in each channel throughout the entire network. 
%Therefore, if there are $m$ channels in a network, there can be at-most $m$ parallel transmissions in the entire network in a time slot. 
The physical channel assigned to a link in a particular time slot is given by~\cite{Chen:2010:WRM:1855162}, $Ch_p = (ASN + Ch_l) \; \mathrm{mod} \; m$, where $ASN$ represents Absolute Slot Number and increases at every slot, $Ch_l$ and $Ch_p$ are the logical and physical channels assigned to a node, $m$ denotes the number of channels in the network.
     
A WirelessHART network is represented as a graph $G=(V,E)$, where $V$ is the set of nodes which are the sensors, actuators and Gateway; $E$ is the set of edges or links between the devices. 
%A node $v \in V$, can be either a sensor, an actuator or a Gateway. 
An edge $e = u \rightarrow v$, $u,v \in V$, is part of $G$, if and only if device $u$ can reliably communicate with device $v$. In a transmission along an edge $u \rightarrow v$, the transmitting node, $u$, is the {\em sender} and the receiving node, $v$, is the {\em receiver} of the transmission.      
%The conflict transmissions/edges and conflict graph are defined as follows.
\begin{defn}
	Two transmissions along edges $u \rightarrow v$ and $w\rightarrow x$, where $u,v,w,x \in V$, are said to be \textbf{conflicting transmissions}, if both of them have the same sender or the same receiver, i.e., if $(u = w) \lor (v = w) \lor (u = x) \lor (v = x)$. For each edge $u \rightarrow v \in E$, there exists a set of conflicting transmissions in $G$. To keep track of the conflicting transmissions in $G$, we store an adjacency list known as the \textbf{Conflict List}. Each index $i$ in the list corresponds to an edge in $E$ and the list corresponding to $i$ stores the list of edges which generate conflicting transmissions with $i$. 
\end{defn}
%\begin{defn}
%	To mark the possibility of conflicting transmissions in a graph $G = (V,E)$, we generate another graph, $G'=(V,E')$, known as the \textbf{Conflict Graph}. For every edge $u \rightarrow v \in E$, $u,v \in V$, there exists a set of conflicting transmissions. 
%	
%	
%	where there is an undirected edge between $u$ and $v$, $u,v \in V$, if $u$ and $v$ does not have an edge in $E$ and $u$ and $v$ results in {\em conflicting transmission} in $G$. All the edges in $E'$ are known as \textbf{Conflict Edges}.	
%\end{defn}
%A WirelesHART network graph consists of an uplink graph $(U)$ (connecting the sensor devices to the Gateway) and a downlink graph $(D)$ (connecting the Gateway to the actuators). 
An end-to-end communication between a sensor and an actuator occurs in two phases: a {\em sensing phase} and a {\em control phase} during which the communications are between the sensors and the Gateway and between the Gateway and the actuators respectively.
\section{System Model}\label{sec:system}

Our system model consists of a WirelessHART network $G=(V,E)$ and $n$ end-to-end flows $\mathcal{F} = \{\mathcal{F}_1, \mathcal{F}_2, \ldots \mathcal{F}_n\}$. 
%Each end-to-end communication in a flow occurs in two phases, a {\em sensing} phase along the uplink graph~$(U)$ and a {\em control} phase along the downlink graph~$(D)$. %Hence, a flow $\mathcal{F}_i \in \mathcal{F}$ can be subdivided into two sub flows, $\mathcal{F}_i^{U}$ and $\mathcal{F}_i^{D}$ along the uplink and downlink graph respectively. 
%A flow in $\mathcal{F}$ is represented as $\mathcal{F}_i = \{s_i, d_i, \delta_i, p_i, R_i\}$. 
Each flow $\mathcal{F}_i \in \mathcal{F}$ periodically generates a packet at the source node $s_i \in V$ with period $p_i$. The packet passes via Gateway and reaches the destination node $d_i \in V \setminus \{s_i\}$ within deadline $\delta_i$. We assume that our flows are of implicit deadline, {\em i.e.}, $\delta_i \leq p_i$. A packet is scheduled in more than one routes between the source and destination for reliability. 
%We use the following definitions.
%where $s_i \in V$ is the source node or a sensor device where the flow originates, $d_i$ is the destination node or an actuator, $\delta_i$ denotes the deadline within which the flow $\mathcal{F}_i$ has to be completed. $p_i$ is the period after which the next flow instance of $\mathcal{F}_i$ is generated at the $s_i$. It is to be noted that $\delta_i \leq p_i$. $R_i$ consists of the set of all routes from $s_i$ to the $d_i$. 
%A flow $\mathcal{F}_i$ for a downlink graph~$(D)$ can also be represented in a similar way.
%as  $\mathcal{F}_i^D = \{s_i^D, d_i^D, \delta_i^D, p_i^D, R_i^D \}$, where $s_i^D$ is the Gateway, $d_i^D$ are the destination nodes or the actuators, $\delta_i^D$ is the deadline within which the control commands are to be delivered from the Gateway to the actuators, $p_i^D$ is the period at which control commands are generated at the Gateway for $d_i^D$ and $R_i^D$ represents the set of all the routes from the Gateway to the actuators. Each flow in a network consists of a number of instances. 
\begin{defn}
	The \textbf{release time} ($r_{ij}$) of the $j^{th}$ instance of flow $\mathcal{F}_i$ ($j \geq  1$) is the time at which the $j^{th}$ instance of $\mathcal{F}_i$ is released at the source node $s_i$. $r_{ij}$  is defined as
	\begin{equation}\label{eq:release_time}
	\scriptsize
	r_{ij} = (j-1) \cdot p_i.
	\end{equation}
\end{defn}  
\begin{defn}
	The \textbf{number of hops} in a route of a flow $\mathcal{F}_i$ is the number of intermediate devices between the source $(s_i)$ and the destination $(d_i)$ in the route of $\mathcal{F}_i$.
\end{defn}
\begin{defn}\label{def:sched}
	Given a graph $G$ with $m$ channels and a set of flows $\mathcal{F}$, a \textbf{feasible schedule}~$\mathcal{S}$ is a sequence of transmissions over the slots in $\mathcal{S}$ along the edges in $G$. Each transmission is a mapping of a flow to a channel in a slot
	% such that each flow $\mathcal{F}_j$, ($1 \leq j \leq |\mathcal{F}|$), is associated with a channel $k$, ($1 \leq k \leq m$) in slot $i$, ($1 \leq i \leq |\mathcal{S}|$) 
	satisfying the following conditions:
	
	\vspace{0.5em}\noindent \textbf{1. No transmission conflict:} Two transmissions along $u \rightarrow v$ and $w \rightarrow x$ can be scheduled in the same time slot $t$, if $u \rightarrow v$ and $w \rightarrow x$ are non-conflicting transmissions;
	
	\vspace{0.5em}\noindent \textbf{2. No collision:} If $u \rightarrow v$ uses channel $y$ and $w \rightarrow x$ uses channel $z$ in the same time slot $t$, then $y \neq z$, $\forall y, z \in [1,m]$;
	
	\vspace{0.5em}\noindent \textbf{3. No deadline violation:} If a flow $\mathcal{F}_j$, $1 \leq j \leq n$, has $h$ hops, then all the $h$ hops of $\mathcal{F}_j$ are to be scheduled within the deadline $\delta_j$;
	
	\vspace{0.5em}\noindent \textbf{4. Flow sequence preservation:} If a flow $\mathcal{F}_j$ has $h$ hops, then the $k^{th}$ hop $(1 < k \leq h)$ cannot be scheduled until all the previous $k-1$ hops are scheduled.	
\end{defn}

We assume that the network manager blacklists those channels from the network in which the probability of successful transmission is less than a certain threshold~\cite{song2008wirelesshart}. Therefore, the number of packet drops in the network can be neglected. At the time of network initialization, the network manager decides the {\em schedule} depending on the number of available channels, the topology of the network and available routes for each flow~\cite{Chen:2010:WRM:1855162},\cite{song2008complete}. Given a graph $G$, a set of $n$ flows $\mathcal{F}$ over $G$ and $m$ channels, the network manager runs any scheduling algorithm~$\mathbb{A}$ that generates a schedule $\mathcal{S}$ satisfying all the conditions of Definition~\ref{def:sched}. 
%following the conditions of scheduling in WirelessHART networks~\cite{han2011reliable}. 
%We assume that there is no priority constraint among the flows in $\mathcal{F}$.
%, i.e., each flow in $\mathcal{F}$ is equally likely to occur as long as all the flows in $\mathcal{F}$ are schedulable. 
The network manager then informs all the network devices about the allocated slots in which they can transmit (receive) messages from specific neighbors. The network devices become active only in those slots in which they can transmit (receive) messages. The same schedule repeats every hyper-period. 
%This makes the schedule predictable for an attacker.
%This paves the way for an attacker to launch some timing attacks, such as the side-channel attacks~(SCA)~\cite{pongaliur2008securing} or selective jamming attacks~\cite{proano2010selective}. 
%We consider the schedule $\mathcal{S}$ over $\mathcal{F}$ over $\mathcal{F}$ generated by $\mathbb{A}$ for our network graph $G$ with $m$ channels such that $\mathcal{S}$ follows all the conditions for scheduling in WirelessHART network~\cite{han2011reliable}. Our objective is to obfuscate $\mathcal{S}$ to  the inherent determinism in $\mathcal{S}$ 
%We consider a schedule $\mathcal{S}$ over $\mathcal{F}$ which is schedulable under $\mathbb{A}$. We assume that there is no  
%
%We assume that there is no priority constraint among the flows in $\mathcal{F}$, i.e., two flows $\mathcal{F}_i$ and $\mathcal{F}_j$ are equally likely to occur in a schedule  ,.
%  
%    
%    
%    All the flows in our network have equal priorities.\label{ass7}
%    During network initialization, the network manager blacklists those channels in which the probability of successful transmission is less then a certain threshold, generally ($80\%- 85\%$). Therefore, the number of message drops over a hyper-period of time is negligibly small.\label{ass8}
\section{Threat Model}\label{sec:threat}

The main objective of the adversary is to select a critical sensor or an actuator as the victim node in the network and predict the time slots in which the victim node sends (receives) packets to (from) its neighboring nodes by observing the traffic in the network. Our adversary model is based on the following assumptions:-
\begin{enumerate}
	\item The adversary is aware of the network parameters such as the number of channels adopted by the network.
	\item The adversary is equipped with multiple antennae, hence, he is capable of listening to all 16 channels in 2.4 GHz ISM band in the network.
	%\item The adversary  has moderate computational capability.
\end{enumerate}
Based on the above assumptions, the adversary has the following capabilities:

\vspace{0.5em}\noindent \textbf{Capability 1:} The adversary can target a specific node (sensor or actuator) as the victim node in the network and monitor all communications associated with that node. After analyzing the traffic for a sufficiently long period of time, the adversary can predict the time slots in which the victim node communicates with its neighbors.
	
\vspace{0.5em}\noindent \textbf{Capability 2:} Due to repetitive nature of the communication schedule, the adversary can estimate the hyper-period of the schedule. The adversary can use this estimate in the subsequent hyper-periods to infer the communication time slots of the victim node.

\vspace{0.5em}\noindent \textbf{Capability 3:} The adversary can reverse engineer the channel hopping sequences by silently observing the channel activities in the network~\cite{mosha2019iotdi}.

% Capability 3- The attacker can launch some timing attacks into the system such as selective jamming attacks by jamming the targeted communication link in a specific time slot in the schedule.

With the above three capabilities, the adversary can execute further destructive attack steps. For instance, 
%using Capability~1 and 2, the adversary has a knowledge of the time slots in which a victim node communicates with its neighbors in a hyper-period schedule. Using Capability~3, the adversary can guess the channel through which the victim node communicates with a specific neighbor in a specific time slot. Thereafter, 
the adversary can target specific transmissions from (to) certain critical sensors (actuators) and can selectively jam the targeted transmissions in specific time slots, thereby causing disruptive effect on the system. Due to repetitive nature of the hyper-period schedules, same flow gets transmitted in the same time slot over every hyper-period. Hence, selectively jamming the predicted channel in specific time slots over every hyper-period results in jamming the targeted flow with probability~$1$. Different from the constant jamming attack that jams all the transmissions, selective jamming is more stealthy as it allows the attacker to strategically target certain critical sensors and/or actuators within their proximity with much lower radio transmission power. This reduces the overhead and cost for the attacker to implement the jamming attack~\cite{grover2014jamming}. In contrast, random jamming that does not infer the schedule and jams in randomly selected slots is much less effective~\cite{xu2005feasibility}.
%For instance, based on the knowledge of when a victim node communicates with its neighbors, the adversary can launch {\em selective jamming} attacks targeting specific transmissions. 

\vspace{0.5em}\noindent \textbf{Attack consequences:} Selectively jamming the transmissions from a critical sensor node results in blocking the sensor data to reach the Gateway. As a result, proper control commands cannot be delivered to the actuators which in turn may result in degraded performance of the system. Also, selectively jamming the control commands to reach the actuators may hamper the safety of the system.

\vspace{0.5em}\noindent \textbf{Motivation of our work: }\label{subsec:intrusion}
The main objective of our work is to develop a MTD technique, the {\em SlotSwapper}, that randomizes the communication time slots over every hyper-period schedule such that the schedule changes before the attacker can estimate it. We present a motivating example to illustrate how the threat can be addressed by randomizing the time slots in every hyper-period schedule.
\begin{figure}
		\includegraphics[width=1.8cm]{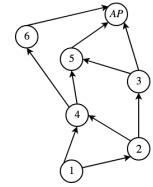}
		\caption{A network graph with six nodes and one AP}
		\label{fig:graph}
\end{figure}
\begin{table}
	\scalebox{0.7}{
		\begin{tabular}{|cccccccccc|} \hline
			&&1 & 2 & 3 & 4 & 5 & 6 & 7 & 8 \\ \hline 
			$\mathcal{S}_1$&&$1 - 2$ & ---  &---&$2 - 3$&$4 - 5$&--- &---& $3 - AP$  \\ 
			&$ch_1$&($F_1$) & & &($F_3$) &($F_2$) & & &($F_3$) \\
			&&$4 - 5$ &--- &$5 - AP$ &--- &$2 - 3$&$3 - AP$ &$5 - AP$ &--- \\
			&$ch_2$&($F_2$) & &($F_2$) & &($F_1$) &($F_1$)&($F_2$) & \\
			\hline
			$\mathcal{S}_2$&&---&$1-2$&---&$5 - AP$&$3-AP$&$4 - 5$&---&$5 - AP$\\
			&$ch_1$&&($F_1$)&&($F_2$)&($F_1$)&($F_2$)&&($F_2$)\\
			&&$2 - 3$&$4 - 5$&$2-3$&---&---&&$3-AP$&---\\ 
			&$ch_2$&($F_3$)&($F_2$)&($F_1$)&&&&($F_3$)&\\ \hline
	\end{tabular}}
	\caption{Two schedules~$\mathcal{S}_1$ and $\mathcal{S}_2$ over $8$ time slots with three flows $\mathcal{F}_1$, $\mathcal{F}_2$, $\mathcal{F}_3$ where $s_1 = 1$, $s_2 = 4$, $s_3 = 2$ and $d_1 = d_2 = d_3 = AP$.}
	\label{table:sched_ad}
\end{table} 
\begin{eg}\label{eg:attack}
	Consider the network graph shown in Figure~\ref{fig:graph} with two channels, three flows, $F_1$, $F_2$ and $F_3$ where the sources are $s_1 = 1$, $s_2 = 4$, $s_3 = 2$; the destinations are $d_1 = d_2 = d_3 = AP$; the periods and the dealines are $p_1 = p_3 = \delta_1 = \delta_3 = 8$, $p_2 = \delta_2 = 4$ respectively. Consider $\mathcal{S}_1$ in Table~\ref{table:sched_ad} to be the hyper-period schedule over the flows. Consider node~$1$ to be the victim node. In the traditional TDMA-based real-time WirelessHART network, the network starts with schedule $\mathcal{S}_1$ which repeats every $8$ time slots. An attacker listening to the channels in the network will find nodes $1$ and $2$ communicating every $8$ time slots. In particular, to identify this repetitive pattern, the attacker needs to listen to the network for at least two hyper-periods, i.e., $16$ time slots. The attacker can launch selective jamming attack earliest in the $17^{th}$ slot. With our proposed MTD technique, a new schedule is followed in each hyper-period, {\em i.e.}, if $\mathcal{S}_1$ is followed in the first eight slots, then $\mathcal{S}_2$ will be followed in the next eight slots and so on. However, there is no communication between nodes $1$ and $2$ in slot~1 in $\mathcal{S}_2$, {\em i.e.}, the communicating time slots in two consecutive hyper-periods are different. To identify the repetitive patterns in the schedule, the attacker needs to monitor the communications for at least two hyper-periods. Hence, by changing the schedule every hyper-period, the system will change at a faster pace compared to the learning pace of the attacker, rendering further strategic destructive attack steps (e.g., selective jamming) infeasible.
\end{eg}
\section{Proposed MTD technique}\label{sec:MTD}
Our proposed MTD technique, the {\em SlotSwapper}, consists of two main phases--- (1) An offline schedule generation phase (2) an online schedule selection phase. $Sched\_Gen()$ considers an initial hyper-period schedule $\mathcal{B}$ for a set of $n$ flows $\mathcal{F}$ over a graph $G$, and generates a new feasible schedule $\mathcal{S}'$ by randomizing the slots in $\mathcal{B}$. However, randomization of time slots in $\mathcal{B}$ is to be done in such a way that all the conditions of generating a {\em feasible schedule} (Definition~\ref{def:sched}) are obeyed. To reduce the repeatability of time slots in $\mathcal{B}$, we propose to run $Sched\_Gen()$ $K$ times ($K$ is a large number) in offline mode and generate a set of feasible hyper-period schedules $\mathbb{S}$. We suggest to select a schedule uniformly at random every hyper-period from $\mathbb{S}$ and execute that schedule over that hyper-period.
\begin{algorithm}
	\scriptsize
	$\mathbb{S}$ = $\{\mathcal{B}\}$;\tcp{a base scehdule}
	\For{i=1,2 upto K}
	{
		 $\mathbb{S}$ = $\mathbb{S} ~\cup$ \textit{$Sched\_Gen()$}\;
	}
	$\mathcal{S}$ = Select a random schedule from $\mathbb{S}$ every hyper-period \;
	\caption{\textit{SlotSwapper}}
	\label{algo:slotswap}
\end{algorithm}
%We consider an initial hyper-period schedule or a base schedule $\mathcal{S}$ for a set of flows $\mathcal{F}$ over a graph $G$ and randomize the slots in $\mathcal{S}$ to generate a new feasible schedule $\mathcal{S}'$.  

\vspace{0.5em}\noindent \textbf{Offline Randomized Schedule Generator :} Algorithm~\ref{algo:swapmultichan} presents an overview of $Sched\_Gen()$. Table~\ref{tab:2} summarizes the notations used in the algorithm. We present an example to illustrate the steps of $Sched\_Gen()$.
%The algorithm has some pre-processing steps as follows-
%\vspace{0.5em}\noindent 1. Make a copy of the base schedule $\mathcal{B}$ in $\mathcal{S}'$ so that all updates are done in $\mathcal{S}'$.
%\vspace{0.5em}\noindent 2. Create two lists of dictionaries $hop\_list$ and $edge\_list$. Each entry in $edge\_list$ is a dictionary that maps the channel to edge in a particular slot in $\mathcal{S}'$. $hop\_list$ contains the hop number to slot mapping of all the flow instances in $\mathcal{F}$.
%	where the length of each of the lists is equal to hyper-period. Each entry in the list is a dictionary that maps the flow to hop number and channel to edge in a particular slot in $\mathcal{S}'$.  
	%	$flow\_list$ and 
	%	This step is n ot required for single-channel network ($m = 1$).
	%	, $flow\_list$ and $hop\_list$, stores the mapping of flow $\mathcal{F}_j$ 
	%	are mappings from flow to channel assignments and flow to hop number,respectively, corresponding to the schedule $\mathcal{S}'$.
\begin{table}[h]
	\scalebox{0.75}{
	\captionsetup{justification=centering}
	\begin{tabular}{|ll|}\hline
		$G$ & a network graph over $V$ nodes and $|E|$ edges\\
		$\mathcal{F}$ & a set of $n$ flows defined over $G$ \\
		$m$ & number of channels in the network \\
		$hp$ & hyper-period of $n$ flows \\
		$\mathcal{B}$ & a base schedule consisting of mapping of a channel in a slot to a \\
		&flow over one $hp$\\ 
%		$flow\_list$ & list of $n$ flows, each entry in the list is a structure that stores the \\ & periods, deadlines, source, destination and routes of a flow\\
		$C$ & Conflict List corresponding to the network graph $G$ \\ 
		$\mathcal{S}'$ & a copy of the base schedule $\mathcal{B}$\\ 
		$hop\_list$ & a dictionary to store hop number to slot mapping of all the\\
		& flow instances in $\mathcal{F}$\\
		$edge\_list$ & a dictionary to map channel to edge in a particular slot in $\mathcal{S}'$.\\
		\hline
	\end{tabular}}
	\caption{List of notations used in the algorithm.}
	\label{tab:2}
\end{table}
\begin{algorithm}[t]
	\scriptsize
	%\DontPrintSemicolon % Some LaTeX compilers require you to use \dontprintsemicolon instead
	%\KwIn{$G$, $\mathcal{F} = \{\mathcal{F}_1,\mathcal{F}_2, \ldots \mathcal{F}_n\}$, $hp$, $m$, $\mathcal{S}$, $Fseq$, $CL$}
	%\KwOut{a new {\em feasible schedule} $\mathcal{S}'$ over $\mathcal{F}$ and $m$ channels}
%	\text{Keep a copy of the base schedule $\mathcal{S}$ in $\mathcal{S}'$}\;
%	\text{Preprocess} 
%	$(hop\_list,flow\_list) = Preprocess\_flow(\mathcal{S}',hp)$\;
	\For{tick = 1,2, \ldots, hp}	
	{
		\For{j = 1,2, \ldots, $|\mathcal{F}|$}
		{
			\If{tick == $\mathcal{F}_j.deadline$}
			{
				$inst = tick/\mathcal{F}_j.deadline$\;
				\For{p = 1,2, \ldots ,$\mathcal{F}_j.n\_hops$}
				{
					$\sigma_t$ = slot of $p^{th}$ hop of $inst$\;
					$elig\_list$ = $\{\}$;\tcp{empty list}
					\If{m == 1 \tcp{single-channel}}
					{
						lb = $inst * \mathcal{F}_j.release\_time$\;
						ub = $inst * \mathcal{F}_j.deadline$\;
						\For{$\sigma_t'$ = lb, lb+1,\ldots, ub}
						{
							\If{$\mathcal{S}'[\sigma_t'] \neq \mathcal{F}_j$}
							{
								Add $\sigma_t'$ to the $elig\_list$\;
%							$elig\_slots$ = $\{s_t'~|~\mathcal{S}'[s_t']~=~\text{idle}~\text{or}~\mathcal{S}'[s_t']~=~\mathcal{F}_j', j~\neq~j'\}$\;
						}
						}
						$\sigma_{random}$ = random($elig\_slots$)\;
						swap ($\sigma_t,\sigma_{random}$)\; 
					}
					\Else
					{		
						$c\_ch$ = channel of $p^{th}$ hop of $inst$\;			
						\If{p == 1\tcp{first hop}}
						{
							lb = $inst$ * $\mathcal{F}_j.period$;
						}
						\Else
						{
							lb =  slot of $(p-1)^{th}$ hop of $inst$ + 1\;
						}
						\If{p == $\mathcal{F}_j.n\_hops$  \tcp{last hop}}
						{
							ub = $inst$ * $\mathcal{F}_j.deadline$\;
						}
						\Else
						{
							ub = slot of $(p+1)^{th}$ hop of $inst$ - 1\;
						}
%						\tcp{slot-channel pair list}
%						$elig\_cand$ = $\{\}$\;
						\For{$\sigma_t'$ = lb, lb+1,\ldots, ub}
						{
							\For{ch = 1,2,\ldots,m}
							{
								$b_1$ = $trConf(\sigma_t,c\_ch,\sigma_t',ch,C)$\;
								$b_2$ = $deadPr(\sigma_t,c\_ch,\sigma_t',ch)$\;
								$b_3$ = $flowPr(\sigma_t,c\_ch,\sigma_t',ch)$\;
								\If{$b_1~\&\&~b_2~\&\&~b3 == 1$}
								{
									Add ($\sigma_t',ch$) to $elig\_list$\;
%								$elig\_cand = \{(s_t',ch)~|~b_1~\&\&~b_2~\&\&~b_3==1\}$\;
								}
							}
						}
						$(\sigma,c)$ = random($elig\_list$)\;
						swap($\sigma_t,c\_ch,\sigma,c$)\;		
					}
					update $hop\_list$, $edge\_list$ and $\mathcal{S}'$\;
				}	
			}
		}	
	}
	\Return $\mathcal{S}'$\;
	\caption{\textit{$Sched\_Gen$}}
	\label{algo:swapmultichan}
\end{algorithm}

%\begin{algorithm}[t]
%	$e_1$ = $edge\_list[s_t].get(c\_ch)$\;
%	$e_2$ = $edge\_list[s_t'].get(ch)$\;
%	\For{i = 1,2,\ldots upto m \&\& i $\neq ch$}	
%	{
%		$e_3$ = $edge\_list[s_t'].get(i)$\;
%		\If{$e_1 \in C[e_3]$ \tcp{$e_1$ is in Conflict List of $e_3$}}
%		{
%			return $0$\;
%		}	
%	}
%	\For{i = 1,2,\ldots upto m \&\& i $\neq c\_ch$}	
%	{
%		$e_3$ = $edge\_list[s_t].get(i)$\;
%		\If{$e_2 \in C[e_3]$ \tcp{$e_2$ is in Conflict List of $e_3$}}
%		{
%			return $0$\;
%		}	
%	}
%	\Return $1$\;	
%	\caption{trConf($(s_t,c\_ch,s_t',ch,C)$}
%	\label{algo:transmission}
%\end{algorithm}
\begin{eg}
	Consider the same setting as in Figure~\ref{fig:graph} and Example~\ref{eg:attack}. Let $\mathcal{S}_1$ in Table~\ref{table:sched_ad} be the base schedule. Let us consider the $1^{st}$ hop of $\mathcal{F}_3$ in $\mathcal{S}_1$ with $\sigma_t = 4$ and $c\_ch = 1$. The window corresponding to $1^{st}$ hop of $\mathcal{F}_3$ is $[1,7]$. For every slot $\sigma_t' \in [1,7]$ and every channel $ch \in [1,2]$, we call $trConf()$ and check for conflicting transmission. $2 \rightarrow 3$ has conflicting transmission with $[1\rightarrow 2, 2\rightarrow 4, 3\rightarrow AP]$ in $\mathcal{S}_1$. Therefore, (slot,channel) pairs such as, $(1,1)$, $(5,2)$ and $(6,2)$ are rejected due to transmission conflict with $(4,1)$. Similarly, (slot,channel) pairs such as $(5,1)$ and $(7,2)$ are also rejected by function $deadPr()$ due to violation of deadlines of the flow instances. (slot,channel) pairs $(5,1)$ and $(7,2)$ correspond to the second instance of $\mathcal{F}_2$ with release time at $5^{th}$ slot and deadline at $8^{th}$ slot. Hence, the second instance of $\mathcal{F}_2$ cannot be swapped with any other slot before slot $5$ or after slot $8$. Similarly, $flowPr()$ does not allow (slot,channel) pairs $(1,2)$, $(6,2)$ and $(7,2)$ in the eligible list in order to preserve the hop sequences of flows. If the transmission corresponding to $1^{st}$ hop of $1^{st}$ instance of $\mathcal{F}_2$ (via edge $4\rightarrow 5$) of (slot,channel) pair $(1,2)$ is allowed to swap with $(4,1)$, then the second hop of that instance of $\mathcal{F}_2$ would have been scheduled before the first hop, violating the hop sequences of the flow instances. Finally, the list of eligible (slot,channel) pairs are --- $[(2,1),(2,2),(3,1),(3,2),(4,2),\\(6,1),(7,1)]$. Let $(3,2)$ be the randomly selected element. Swapping the transmissions and the flow instances between $(3,2)$ and $(4,1)$ and iterating the same procedure over all the flow instances generates a completely new feasible schedule. 
%	Selecting a new schedule in every hyper-period makes the schedule completely unpredictable by an attacker and makes the system resilient to timing attacks such as the selective jamming attacks.	 
\end{eg}
\vspace{0.5em}\noindent \textbf{Online Selection of Schedules: }\label{subsec:selection} On executing $Sched\_Gen()$ $K$ times in offline mode, we get a set of feasible schedules $\mathbb{S}$. At the time of network initialization, each node is informed about the time-slots in which it can send/receive messages in each of these $K$ hyper-period schedules. The online schedule selector runs at each node once in every hyper-period, selects a schedule $\mathcal{S}$ from $\mathbb{S}$ uniformly at random and executes $\mathcal{S}$ over that hyper-period. To ensure that the same schedule is selected at each node, we propose to use a pseudo-random number generator (PRNG)~\cite{wiki:prng} (assumed to be secure) initialized with the same seed at each node. This allows each node to select the same schedule every hyper-period without any additional communication. 
\section{Measure of Uncertainty}\label{sec:metric}

%Given a base schedule $\mathcal{S}$ of length $l$ and a set of $n$ flows $\mathcal{F}$, the {\em SlotSwapper} algorithm discussed in Section~\ref{sec:MTD} generates a new feasible schedule $\mathcal{S}'$ by randomizing the slots in $\mathcal{S}$ so that the determinism in the slots in $\mathcal{S}$ is obfuscated in the new schedule $\mathcal{S}'$. The main objective is to make the schedule dynamic so that the predictability in the time slots is reduced in every hyper-period. 
Given a set of schedules $\mathbb{S}$ generated by $Sched\_Gen()$, we need to quantify the amount of uncertainty in the schedules in $\mathbb{S}$. In \cite{yoon2016taskshuffler}, {\em schedule entropy} is used to measure the uncertainty of a given schedule for a uniprocessor system. We have redefined {\em schedule entropy} as a function of the slot and channel entropy to measure the randomness in the schedules in $\mathbb{S}$. In a multi-channel WirelessHART network, each of the slots $\sigma_i$ in a schedule $\mathbb{S}$ consists of $m$ channels which can be represented as $\sigma_i = \{c_{i1},c_{i2},\ldots,c_{im}\}$. Given a hyper-period schedule $\mathcal{S}$ over $l$ slots and $m$ channels for a set of flows $\mathcal{F}$, the occurrence of the $j^{th}$ flow $\mathcal{F}_j$ in the $k^{th}$ channel of $i^{th}$ slot is a discrete random variable with possible outcomes from $0$ to $n$, where $0$ represents idle flow, $n$ is the total number of flows in $\mathcal{F}$. Let $c_{ik} = j$ denotes the $j^{th}$ flow occurring in the $k^{th}$ channel of $i^{th}$ slot of $\mathcal{S}$. However, the occurrence of the $j^{th}$ flow in the $k^{th}$ channel of the $i^{th}$ slot restricts the occurrence of some other flow $\mathcal{F}_j'$ in the same channel of the same slot. Also, if a flow $\mathcal{F}_j$ completes its hops in the $i^{th}$ slot in the schedule, it cannot occur in the subsequent slots until the arrival of its next instance. We therefore, define {\em Schedule entropy} as 
\begin{defn}
	\textbf{Schedule entropy} over a set of flows~$\mathcal{F}$ for a WirelessHART network with $m$ channels is the conditional entropy of $\mathcal{F}_j$ occurring in the $k^{th}$ channel of the $i^{th}$ slot, given the entropy of all the slots from $1$ to $i-1$. It is represented as
	\begin{align}\label{ent}
	\scriptsize
	H(\mathcal{S})=\sum_{i=1}^l H(\sigma_i|\sigma_1,\sigma_2,&\ldots,\sigma_{i-1})\\
%	H(s_1) = - \sum_{j_1=0}^n \sum_{j_2=0}^n \ldots \sum_{j_m=0}^n \Pr(j_1,j_2,\ldots,j_m) &\nonumber\\\log_2 \Pr(j_1,j_2,\ldots,j_m)&\\
	H(\sigma_i)=-\sum_{c_{i1}=0}^n\sum_{c_{i2}=0}^n\ldots&\sum_{c_{im}=0}^n \Pr(c_{i1},c_{i2},\ldots,c_{im})\nonumber\\
	&\log_2 \Pr(c_{i1},c_{i2},\ldots,c_{im})
%	H(\mathcal{S})=\sum_{i = 1}^{l} H((ch_{i1}ch_{i2}\ldots ch_{ik})|(ch_{11} ch_{12}\ldots ch_{1k}) &\\
%	(ch_{21}ch_{22}\ldots ch_{2k}), \ldots (ch_{(i-1)1}ch_{(i-1)2} \ldots ch_{(i-1)k})),& \\
%	H(s_i|s_1\ldots,s_{i-1}) \!=\! \sum_{j_1=1}^n\ldots \sum_{j_i=1}^n \Pr&(s_1=j_1,\ldots,s_i=j_i)\nonumber\\ \log_2&\frac{\Pr(s_i=j_i)}{\Pr(s_1=j_1,\ldots,s_i=j_i)}.
	\end{align}
\end{defn}
For a multi-channel WirelessHART network with $n$ flows ($n > 16$), the number of possible permutations in the calculation of the joint probability for each slot is exponential. Hence, we consider the empirical probability distribution of the flows across all the channels in each slot which is an upper-approximated value of {\em slot entropy} as the joint probability is always less than or equal to the sum of individual probabilities~\cite{shannon2001mathematical}. Further, calculation of conditional entropy in Equation~\eqref{ent} involves joint probability distribution of slots in $\mathcal{S}$, which is exponential in nature. So, we consider the empirical probability distribution of the slots in $\mathcal{S}$. 
\begin{defn}
	\textbf{Upper-approximated slot entropy} $\widetilde{H(\sigma_i)}$  and \textbf{Upper-approximated schedule entropy} $\widetilde{H(\mathcal{S})}$ are defined respectively as follows
	\begin{align}
	\scriptsize
	\widetilde{H(\sigma_i)} &= - \sum_{k = 1}^m \sum_{j = 0}^n \Pr(c_{ik}=j) \log_2 \Pr(c_{ik}=j)\\
	\widetilde{H(\mathcal{S})} &= \sum_{i = 1}^l \widetilde{H(\sigma_i)}.
	\end{align}
	where $\Pr(c_{ik}=j)$ is the probability mass function of the $j^{th}$ flow occurring in the $k^{th}$ channel of the $i^{th}$ slot. 
\end{defn}

\section{Evaluation}\label{sec:eval}
\begin{figure*}
	\begin{minipage}{0.3\linewidth}
		\includegraphics[width=5.7cm]{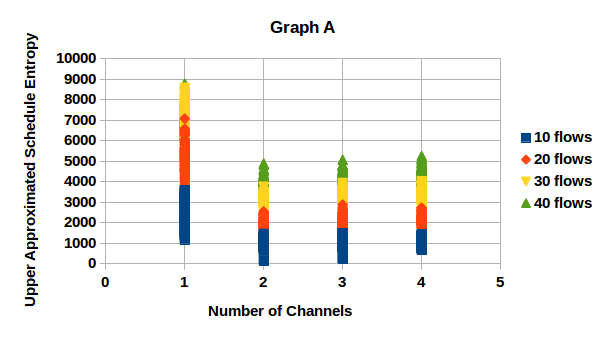}
	\end{minipage}
	~
	\begin{minipage}{0.3\linewidth}
		\includegraphics[width=5.7cm]{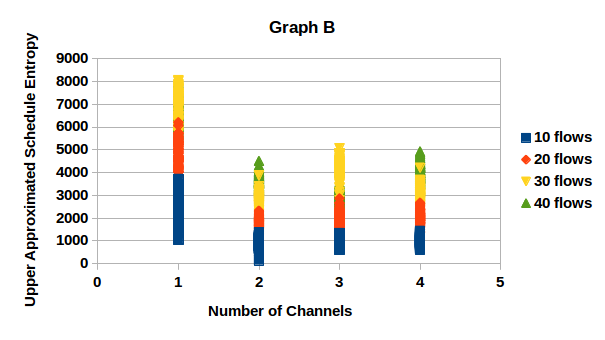}
	\end{minipage}
	~
	\begin{minipage}{0.3\linewidth}
		\includegraphics[width=5.7cm]{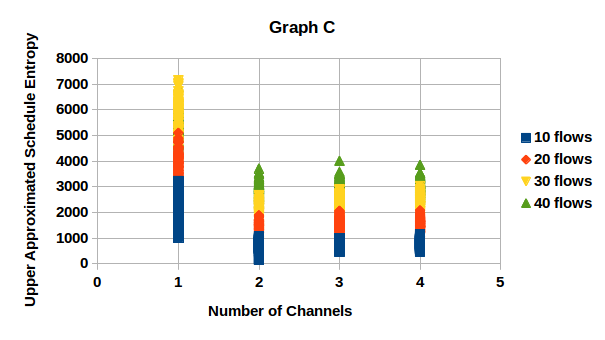}
	\end{minipage}
	\caption{Upper Approximated Schedule Entropy over Graph A,Graph B and Graph C, with number of flows varying between 10 to 40 and number of channels between 1 to 4 with a hyper-period of 1024 time slots}
	\label{fig:entropy}
\end{figure*}
%\vspace{0.5em}\noindent \textbf{Testbed Setup:} We carried out extensive evaluation of our SlotSwapper algorithms on the Indriya2 testbed~\cite{indriya} consisting of 74 TelosB motes. Based on the connectivity and positions of the nodes, we assigned three AP in the network.

\vspace{0.5em}\noindent \textbf{Simulation setup:} We use Cooja simulator~\cite{osterlind2006cross} of Contiki 3.0 to test the feasibility of our schedules. We generated three random topologies with 100 simulated Tmote Sky motes by varying the degree of nodes ($\theta$) or the number of incoming and outgoing edges incident on a node --- (1) Graph A ($\theta$ between 2 to 4) (2) Graph B ($\theta$ between 3 to 6) (3) Graph C ($\theta$ between 3 to 8). More the degree of a node, more are the chances of conflicting transmissions and less is the number of available flows for a particular time-slot. Nodes with highest number of neighbors are considered to be the APs.

\vspace{0.5em}\noindent \textbf{Flow Generation:} A fraction ($\alpha$) percent of the nodes are randomly selected as the source and destination nodes. The source and destination nodes are disjoint. In our experiments we varied $\alpha$ between 20-80\%. 
%For the testbed, the paths of the flows are generated based on the Packet Reception Ratio (PRR) of the links. In the testbed, we were able to generate a maximum of 28 random flows. 
%We selected the shortest path (number of hops) between the source and the destination node as the primary path. 
We selected the number of hops of each flow to be between 2 to 8~\cite{alur2009modeling} and considered the shortest path as the primary path. The flows have implicit-deadline with periods varying randomly in the range of $2^7$ to $2^{10}$.  

\vspace{0.5em}\noindent \textbf{Experiments:} We fixed the hyper-period at $2^{10}$ time slots and ran experiments upto 10000 hyper-periods with the number of flows and the number of channels varying between 10 to 40 and 1 to 4 respectively. For each condition, we generated 100 random instances and measured the upper approximated schedule entropy ($\widetilde{H(\mathcal{S})}$) for each of these instances. Figure \ref{fig:entropy} shows $\widetilde{H(\mathcal{S})}$ for all the tested scenarios. It has been observed that $\widetilde{H(\mathcal{S})}$ is maximum for single-channel WirelessHART network for all three graphs. This is because in single-channel WirelessHART networks, there is no conflicting transmissions among the flows in the network. As a result, a flow can be scheduled at any slot within its release time and deadline. For a fixed number of channels, $\widetilde{H(\mathcal{S})}$ increases significantly with increase in the number of flows upto 30. After that, there is no significant increase in the value of $\widetilde{H(\mathcal{S})}$ with increase in the number of flows. This is because, with increase in the number of flows more flows can appear in a slot. However, as the number of flows increase, the number of conflicting transmissions among the flows increase which in turn restricts the number of available flows to be scheduled in a particular slot. $\widetilde{H(\mathcal{S})}$ also increases with increase in the number of channels between 2 to 4, as the number of available positions for a flow to be scheduled get increased. However, it has been observed that with increase in the number of channels, the increase in $\widetilde{H(\mathcal{S})}$ is significantly less for Graph~C. Among all the three graphs, the number of edges is maximum in Graph~C resulting in more conflicting transmissions among the flows thereby restricting the number of available positions to schedule a flow. 

Although we ran our algorithm upto 10000 hyper-periods to measure the randomness in the generated schedules, the amount of memory available to each Tmote sky mote is not sufficiently large to store large number of schedules. We measured that each mote can only support a maximum of 2000 time slot information. We observed that, if a node is in the path of all the 40 flows, then it requires to store at-least 80 time slot information per schedule (40 for transmissions and 40 for re-transmissions). With this specification, we were able to store 25 schedules in each node. We can manually tune the nodes with different sets of schedules after several hyper-periods to further reduce the chance of predicting the schedules. Our MTD technique only involves an additional random number generation in each node once in every hyper-period, the power consumption of which is negligibly small.

\section{Conclusion}\label{sec:conclusion}

In this work, we presented an MTD mechanism, the {\em SlotSwapper}, to reduce the predictability of TDMA slots in a real-time WirelessHART network. We used schedule entropy to measure the uncertainty of the schedules generated by our algorithm. We illustrated the feasibility of the schedules on simulated networks in Cooja with 100 Tmote sky motes. 

\section{Acknowledgement}

This work was conducted within the Delta-NTU Corporate Lab for Cyber-Physical Systems with funding support from Delta Electronics Inc. and the National Research Foundation (NRF) Singapore under the Corp Lab@University Scheme.

\scriptsize
\bibliographystyle{unsrt}
\bibliography{reference}

\begin{thebibliography}{10}

\bibitem{langner2011stuxnet}
Ralph Langner.
\newblock Stuxnet: Dissecting a cyberwarfare weapon.
\newblock {\em IEEE Security \& Privacy}, 2011.

\bibitem{dragonfly}
Dragonfly: Western energy sector targeted by sophisticated attack group, 2017.
\newblock \url{https://symc.ly/2Df3VTi}.

\bibitem{yoon2016taskshuffler}
Man-Ki Yoon, Sibin Mohan, Chien-Ying Chen, and Lui Sha.
\newblock Taskshuffler: A schedule randomization protocol for obfuscation
  against timing inference attacks in real-time systems.
\newblock In {\em Real-Time and Embedded Technology and Applications Symposium
  (RTAS), 2016 IEEE}. IEEE.

\bibitem{osterlind2006cross}
Fredrik Osterlind, Adam Dunkels, Joakim Eriksson, Niclas Finne, and Thiemo
  Voigt.
\newblock Cross-level sensor network simulation with cooja.
\newblock In {\em Local computer networks, proceedings 2006 31st IEEE
  conference on}. IEEE.

\bibitem{jiang2016sparta}
Ke~Jiang, Petru Eles, Zebo Peng, Sudipta Chattopadhyay, and Lejla Batina.
\newblock Sparta: A scheduling policy for thwarting differential power analysis
  attacks.
\newblock In {\em Design Automation Conference (ASP-DAC), 2016 21st Asia and
  South Pacific}. IEEE.

\bibitem{proano2010selective}
Alejandro Proano and Loukas Lazos.
\newblock Selective jamming attacks in wireless networks.
\newblock In {\em Communications (ICC), 2010 IEEE International Conference on}.
  IEEE.

\bibitem{mpitziopoulos2009survey}
Aristides Mpitziopoulos, Damianos Gavalas, Charalampos Konstantopoulos, and
  Grammati Pantziou.
\newblock A survey on jamming attacks and countermeasures in wsns.
\newblock {\em IEEE Communications Surveys \& Tutorials}, 2009.

\bibitem{virmani2014routing}
Deepali Virmani, Ankita Soni, Shringarica Chandel, and Manas Hemrajani.
\newblock Routing attacks in wireless sensor networks: A survey.
\newblock {\em arXiv preprint arXiv:1407.3987}, 2014.

\bibitem{pongaliur2008securing}
Kanthakumar Pongaliur, Zubin Abraham, Alex~X Liu, Li~Xiao, and Leo Kempel.
\newblock Securing sensor nodes against side channel attacks.
\newblock In {\em High Assurance Systems Engineering Symposium, 2008. HASE
  2008. 11th IEEE}. IEEE.

\bibitem{pickholtz1982theory}
Raymond Pickholtz, Donald Schilling, and Laurence Milstein.
\newblock Theory of spread-spectrum communications--a tutorial.
\newblock {\em IEEE transactions on Communications}, 1982.

\bibitem{proano2012packet}
Alejandro Proano and Loukas Lazos.
\newblock Packet-hiding methods for preventing selective jamming attacks.
\newblock {\em IEEE Transactions on dependable and secure computing}, 2012.

\bibitem{wood2007deejam}
Anthony~D Wood, John~A Stankovic, and Gang Zhou.
\newblock Deejam: Defeating energy-efficient jamming in ieee 802.15. 4-based
  wireless networks.
\newblock In {\em Sensor, Mesh and Ad Hoc Communications and Networks, 2007.
  SECON'07. 4th Annual IEEE Communications Society Conference on}. IEEE.

\bibitem{stojanovski2015efficient}
Spase Stojanovski and Andrea Kulakov.
\newblock Efficient attacks in industrial wireless sensor networks.
\newblock In {\em ICT Innovations 2014}. Springer.

\bibitem{tiloca2017jammy}
Marco Tiloca, Domenico De~Guglielmo, Gianluca Dini, Giuseppe Anastasi, and
  Sajal~K Das.
\newblock Jammy: a distributed and dynamic solution to selective jamming attack
  in tdma wsns.
\newblock {\em IEEE Transactions on Dependable and Secure Computing}, 2017.

\bibitem{tiloca2018dish}
Marco Tiloca, Domenico~De Guglielmo, Gianluca Dini, Giuseppe Anastasi, and
  Sajal~K Das.
\newblock Dish: Distributed shuffling against selective jamming attack in ieee
  802.15. 4e tsch networks.
\newblock {\em ACM Transactions on Sensor Networks (TOSN)}, 2018.

\bibitem{lu2016real}
Chenyang Lu, Abusayeed Saifullah, Bo~Li, Mo~Sha, Humberto Gonzalez, Dolvara
  Gunatilaka, Chengjie Wu, Lanshun Nie, and Yixin Chen.
\newblock Real-time wireless sensor-actuator networks for industrial
  cyber-physical systems.
\newblock {\em Proceedings of the IEEE}, 2016.

\bibitem{Chen:2010:WRM:1855162}
Deji Chen, Mark Nixon, and Aloysius Mok.
\newblock {\em WirelessHART: Real-Time Mesh Network for Industrial Automation}.
\newblock Springer Publishing Company, Incorporated, 2010.

\bibitem{song2008wirelesshart}
Jianping Song, Song Han, Al~Mok, Deji Chen, Mike Lucas, Mark Nixon, and Wally
  Pratt.
\newblock Wirelesshart: Applying wireless technology in real-time industrial
  process control.
\newblock In {\em IEEE real-time and embedded technology and applications
  symposium}. IEEE, 2008.

\bibitem{song2008complete}
Jianping Song, Song Han, Xiuming Zhu, Aloysius~K Mok, Deji Chen, and Mark
  Nixon.
\newblock A complete wirelesshart network.
\newblock In {\em Proceedings of the 6th ACM conference on Embedded network
  sensor systems}. ACM, 2008.

\bibitem{mosha2019iotdi}
Xia Cheng, Junyang Shi, and Mo~Sha.
\newblock Cracking the channel hopping sequences in ieee 802.15.4e-based
  industrial tsch networks.
\newblock 2019.

\bibitem{grover2014jamming}
Kanika Grover, Alvin Lim, and Qing Yang.
\newblock Jamming and anti-jamming techniques in wireless networks: a survey.
\newblock {\em International Journal of Ad Hoc and Ubiquitous Computing}, 2014.

\bibitem{xu2005feasibility}
Wenyuan Xu, Wade Trappe, Yanyong Zhang, and Timothy Wood.
\newblock The feasibility of launching and detecting jamming attacks in
  wireless networks.
\newblock In {\em Proceedings of the 6th ACM international symposium on Mobile
  ad hoc networking and computing}. ACM, 2005.

\bibitem{wiki:prng}
{Wikipedia contributors}.
\newblock Pseudorandom number generator --- {Wikipedia}{,} the free
  encyclopedia, 2019.

\bibitem{shannon2001mathematical}
Claude~Elwood Shannon.
\newblock A mathematical theory of communication.
\newblock {\em ACM SIGMOBILE mobile computing and communications review}, 2001.

\bibitem{alur2009modeling}
Rajeev Alur, Alessandro D'Innocenzo, Karl~H Johansson, George~J Pappas, and
  Gera Weiss.
\newblock Modeling and analysis of multi-hop control networks.
\newblock In {\em Real-Time and Embedded Technology and Applications Symposium,
  2009. RTAS 2009. 15th IEEE}. IEEE, 2009.

\end{thebibliography}
\end{document}